\documentclass[10pt, onecolumn, conference]{IEEEtran}

\usepackage{graphicx,url,tabularx,booktabs}
\usepackage{array}
\usepackage{microtype}
\usepackage[english]{babel}
\usepackage[utf8]{inputenc}  
\usepackage{xspace}

\newcommand{\toolname}{AnonLFI\xspace}
 
\newcommand{\toolurl}{\url{https://github.com/AnonShield/AnonLFI2.0}\xspace} 

\title{AnonLFI 2.0: Extensible Architecture for PII Pseudonymization in CSIRTs with OCR and Technical Recognizers}

\author{
\IEEEauthorblockN{Cristhian Kapelinski, Douglas Lautert, Beatriz Machado, Diego Kreutz}
\IEEEauthorblockA{
AI Horizon Labs -- Federal University of Pampa (UNIPAMPA)\\
\{cristhianavila, douglaslautert, beatrizmachado\}.aluno@unipampa.edu.br, kreutz@unipampa.edu.br
}
}

\begin{document} 

\maketitle

\begin{abstract}
This work presents AnonLFI 2.0, a modular pseudonymization framework for CSIRTs that uses HMAC-SHA256 to generate strong and reversible pseudonyms, preserves XML and JSON structures, and integrates OCR and technical recognizers for PII and security artifacts. In two case studies—OCR on PDF and OpenVAS XML report—the system achieved 100\% precision and F1 scores of 76.5\% and 92.13\%, demonstrating effectiveness in secure preparation of complex cybersecurity datasets.
\end{abstract}


\section{Introduction}
\label{sec:introducao}

Processing incident data by Computer Security Incident Response Teams (CSIRTs) is essential for automated threat analysis, detection engineering, and LLM training \cite{CTISharing}. However, these data frequently contain Personally Identifiable Information (PII), such as IP addresses, credentials, and \textit{hostnames}, whose unrestricted disclosure violates regulations such as GDPR and LGPD. Thus, a structural tension arises between \textit{confidentiality} and \textit{analytical utility}. Classical anonymization approaches, such as k-anonymization \cite{SweeneyKAnon} or textual redaction, are essentially destructive, as they remove the semantic granularity necessary for correlating Indicators of Compromise (IoCs), compromising behavioral analyses, graph-based \textit{threat hunting}, and predictive models.

Pseudonymization constitutes the most suitable strategy for this domain, but existing solutions present critical limitations in three dimensions: (1) cryptographic security, often reduced due to the use of public \textit{hashes} vulnerable to \textit{rainbow table} attacks; (2) structural preservation, with tools that flatten hierarchical formats like XML/JSON or ignore PII embedded in images; and (3) semantic coverage, as many do not recognize fundamental technical entities, such as malware \textit{hashes}, certificate serial numbers, and vulnerability identifiers. The previous version of \toolname \cite{ArtigoAnterior} validated the hybrid approach based on NER/RegEx, but had a monolithic architecture and suffered from these same gaps.

This paper presents \toolname 2.0, a complete reengineering into a modular \textit{framework} designed for cybersecurity requirements at scale. Main contributions include: (1) cryptographically robust pseudonymization via HMAC-SHA256 with secret key, enabling federated correlation and mitigating inversion attacks; (2) a processing \textit{pipeline} composed of native processors capable of preserving the structure of XML and JSON files, plus an OCR module for extracting PII in images and PDF documents\footnote{Partner CSIRTs reported incidents recorded in DOCX and PDF documents containing embedded screenshots.}; (3) specialized technical recognizers, such as HASH and CERT\_SERIAL; and (4) a CLI for controlled re-identification with audit trail support. The empirical evaluation comprises two representative case studies: a PDF processed via OCR (Precision = 100\%, F1 = 76.5\%) and an OpenVAS XML report (Precision = 100\%, F1 = 92.13\%). Results demonstrate that \toolname 2.0 offers secure pseudonymization, preserves analytical utility, and is suitable for building complex \textit{datasets} in cybersecurity.

\section{State of the Art}
\label{sec:relacionados}

The state of the art in PII de-identification ranges from cloud services (e.g., Google DLP, Amazon Comprehend) and open \textit{frameworks} (e.g., Microsoft Presidio) to academic approaches oriented toward preserving semantic utility \cite{Vakili2024, Masketeer2024}. Table~\ref{tab:trabalhos-relacionados} synthesizes this landscape and positions \toolname in relation to the main existing solutions.

While commercial services tend to focus on textual redaction or opaque processing mechanisms, and recent academic proposals \cite{Vakili2024, Masketeer2024} prioritize semantic coherence for model training, \toolname 2.0 emphasizes structural integrity and cryptographic security as central requirements for the cybersecurity domain. Unlike Microsoft Presidio, which demands extensive customization to accommodate specific security workflows, \toolname 2.0 natively provides hierarchical processing of technical reports in XML/JSON, support for technical entity recognition, and a controlled re-identification mechanism based on symmetric keys. These advances overcome the security, extensibility, and formatting limitations identified in version 1.0 \cite{ArtigoAnterior}.

\begin{table}[h!]
\caption{Comparison of de-identification approaches}
\label{tab:trabalhos-relacionados}
\centering
\footnotesize 
\setlength{\tabcolsep}{4pt} 
\begin{tabularx}{\linewidth}{l >{\raggedright\arraybackslash}X >{\raggedright\arraybackslash}X}
\toprule
\textbf{Solution} & \textbf{Data Scope} & \textbf{Main Technique} \\
\midrule
Google DLP & Structured and Text & Hashing (HMAC) or Redaction. \\
Amazon Comprehend  & Text (NLP) & Detection and Redaction (masking). \\
Microsoft Presidio  & Text, Images (OCR) & Extensible framework (NER/RegEx). \\
Vakili et al. (2024) \cite{Vakili2024} & Clinical Text & Replacement with realistic \textit{surrogates}. \\
\midrule
\toolname 1.0 \cite{ArtigoAnterior} & TXT, CSV, DOCX, XML/XLSX (conversion to CSV). & Public SHA256 hash (vulnerable). \\
\textbf{\toolname 2.0} & \textbf{PDF, Images (OCR), XML/JSON (native), TXT, CSV, DOCX.} & \textbf{HMAC-SHA256 (secure) and Reversible.} \\
\bottomrule
\end{tabularx}
\end{table}

\subsection{AnonLFI 1.0}
\label{sec:limitacoes}

AnonLFI 1.0 \cite{ArtigoAnterior} established the first dedicated \textit{pipeline} for pseudonymizing real security incidents processed by Brazilian CSIRTs. Its empirical evaluation, based on a hybrid approach combining spaCy NER and regular expressions, was performed on 763 real incidents and achieved Precision = 100\% and Recall = 97.38\%. Although these results validate the feasibility of the approach, the monolithic architecture of version 1.0 revealed seven structural limitations that restricted security, scalability, and applicability to more complex scenarios:
(L1) the use of \texttt{hashlib.sha256} without a secret key made pseudonyms vulnerable to \textit{rainbow table}-based attacks, compromising confidentiality;
(L2) fixed truncation to 10 characters introduced high collision risk according to the birthday paradox, harming entity correlation;
(L3) the absence of specialized recognizers for technical entities (e.g., malware \textit{hashes}, X.509 certificate serial numbers, CPE strings) limited extensibility to new domains and formats;
(L4) conversion of XML or XLSX to CSV destroyed hierarchy and semantics of structured data, reducing analytical utility;
(L5) there was no support for detecting and pseudonymizing PII present in \textit{screenshots} or image-based documents;
(L6) the reversal process depended on manual SQLite queries, making controlled and auditable re-identification impossible as required by regulatory practices;
(L7) language support was restricted to Portuguese, making multilingual processing common in incident response environments unfeasible.

\section{AnonLFI 2.0: Architecture and Evolution}
\label{sec:arquitetura}

AnonLFI 2.0\footnote{\toolurl} redesigns the previously monolithically implemented \textit{pipeline}, now adopting a modular \textit{framework} composed of four decoupled components: (1) a CLI responsible for routing and handling input files; (2) a central engine based on Microsoft Presidio, in charge of orchestrating language-specific spaCy models and a multilingual Transformer model (\texttt{Davlan/xlm-roberta-base-ner-hrl}); (3) a processing module structured according to the \textit{Factory} pattern, with dedicated processors for each format (PDF, JSON, XML, image, DOCX, XLSX); and (4) a configuration module that externalizes critical execution parameters and pseudonymization policies.

\textit{Recognition engine}: The architecture maintains and extends the original hybrid engine, combining Microsoft Presidio with language-specific spaCy models, the multilingual Transformer model \texttt{Davlan/xlm-roberta-base-ner-hrl}, and specialized regular expressions. Performance metrics vary according to the characteristics of each dataset, considering entity prevalence, syntactic complexity, and application domain. The scenarios evaluated in this work introduce challenges not present in the original 763 incidents, including the need for PII extraction via OCR and recognition of specialized technical entities.

Table~\ref{tab:limitacoes-solucos} presents the mapping between limitations of the previous version and the technical and architectural solutions introduced in AnonLFI 2.0. The proposed solutions systematically address the problems identified in v1.0. In terms of security (L1), the use of HMAC-SHA256 with \texttt{SECRET\_KEY} defined as an environment variable prevents \textit{rainbow table}-based attacks and enables federated correlation between pseudonymized datasets. To ensure integrity and avoid collisions (L2), the \texttt{--slug-length} parameter (default 64) reduces collision probability to negligible values ($< 10^{-60}$), while storing the complete 256-bit hash in the database ensures precise re-identification.

\begin{table}[h!]
\caption{AnonLFI 2.0: solutions to version 1.0 limitations}
\label{tab:limitacoes-solucos}
\centering
\small

\newcolumntype{Y}{>{\hsize=.4\hsize}X}   
\newcolumntype{Z}{>{\hsize=.6\hsize}X}   

\begin{tabularx}{\linewidth}{c Y Z}
\toprule
\textbf{ID} & \textbf{Limitation} & \textbf{Solution} \\
\midrule
L1 & Vulnerable SHA256 & HMAC-SHA256 with SECRET\_KEY \\
L2 & Collisions (10 chars) & Configurable \texttt{--slug-length} (default 64) \\
L3 & No technical recognizers & HASH, CERT\_SERIAL, CERT\_BODY, CPE \\
L4 & CSV flattening & Native JSON/XML/XLSX processors \\
L5 & No images & OCR pipeline (Tesseract) \\
L6 & Manual re-identification & CLI \texttt{deanonymize.py} with audit \\
L7 & Fixed language & \texttt{--lang} parameter (24 languages) \\
\bottomrule
\end{tabularx}
\end{table}

On the extensibility axis (L3), the inclusion of new regular expression-based recognizers (\texttt{HASH}, \texttt{CERT\_SERIAL}, \texttt{CERT\_BODY}, \texttt{CPE\_STRING}) enables processing of technical entities common in vulnerability reports. Regarding structured format handling (L4), the \texttt{JsonFileProcessor} and \texttt{XmlFileProcessor} components recursively traverse native file structures, preserving the original hierarchy. Expanded data coverage (L5) is achieved with the \texttt{ImageFileProcessor}, which applies OCR (Tesseract) to both standalone images and images embedded in documents.

Re-identification (L6) is now conducted by a dedicated CLI (\texttt{deanonymize.py}), which requires the presence of \texttt{SECRET\_KEY} and records audit events. Finally, multilingual support (L7) is made flexible through the \texttt{--lang} parameter, which dynamically loads spaCy models according to the desired language, complemented by the \texttt{--allow-list} parameter, which defines terms that should be preserved.

\section{Case Studies}
\label{sec:estudo-caso}

Two case studies were used to validate the main architectural capabilities introduced in AnonLFI 2.0. The evaluation employed standard NER metrics: Precision ($TP/(TP+FP)$), Recall ($TP/(TP+FN)$), and F1-Score. The \textit{ground truth} was established through manual annotation independently performed by two security researchers, ensuring consistency and reliability of labels.

\subsection{Scenario 1: PDF with Images (OCR)}

\begin{figure}[h!]
    \centering
    \includegraphics[width=0.95\linewidth]{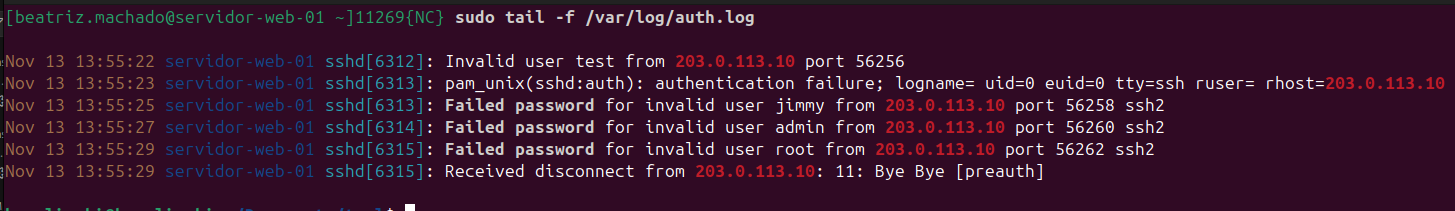}
    \caption{Terminal screenshot with PIIs in high and low contrast areas}
    \label{fig:ssh_log_screenshot}
\end{figure}

This scenario validates the OCR \textit{pipeline} (solution L5) using a security incident report (\texttt{incidente\_ssh.pdf}) concerning an SSH attack. The document includes PII both in digital text and in a terminal screenshot with \textit{syntax highlighting} (Figure~\ref{fig:ssh_log_screenshot}).

\noindent \textbf{Results}: The ground truth contains 21 sensitive entities. TP = 13, FP = 0, and FN = 8 were obtained, resulting in \textbf{Precision = 100\%}, \textbf{Recall = 61.9\%}, and \textbf{F1 = 76.5\%}.

\noindent \textbf{Analysis}: False Negatives have two main origins: (1) \textit{OCR failures (5 cases)}: the \textit{hostname} \texttt{servidor-web-01} was read as \texttt{seryidor web 01} due to low contrast between dark blue and the purple background of the Ubuntu terminal. This type of artifact illustrates inherent OCR limitations, whose accuracy strongly depends on visual quality and image contrast. (2) \textit{Recognition engine failures (3 cases)}: absence of a specific recognizer for the \texttt{[user@host]} pattern in the prompt \texttt{[beatriz.machado@servidor-web-01]}; partial recognition of compound name (inappropriately preserving the surname \texttt{Machado}); and non-classification of the \textit{hostname} when inserted in log context. Among the 13 correctly identified entities, 6 were IP addresses extracted via OCR, demonstrating the method's effectiveness when visual contrast is adequate.

\subsection{Scenario 2: Vulnerability XML (OpenVAS)}

This scenario validates technical recognizers (solution L3) and structural preservation (solution L4). An OpenVAS scan executed against \texttt{localhost} generates a native XML file with nested hierarchical structure, containing specialized technical entities (e.g., \textit{hashes}, X.509 certificates, CPE strings) and a significant volume of structured data.

\begin{verbatim}
$ uv run anon.py vulnerabilidadexml.xml --lang en \
  --slug-length 10 --preserve-entities "CPE_STRING" \
  --allow-list "App,OS,Done,UTC,Default Accounts, \
   Greenbone,Greenbone AG,Greenbone Community Feed, \
   GCF,MQTT Broker Detection, (TCP),Redis"
\end{verbatim}

\textbf{Iterative process}: Initial validation revealed false positives involving product names, metadata, generic technical terms, a numeric timestamp, and the expression \texttt{MQTT Broker Detection (TCP)}, identified by auditing the SQLite database. Applied corrections consisted of refining the \texttt{--allow-list} parameter and adjusting the hexadecimal \textit{hostname} RegEx to exclude \textit{timestamp} patterns.

\noindent \textbf{Results}: The reference set contained 48 sensitive entities. TP = 41, FP = 0, and FN = 7 were obtained, resulting in \textit{Precision = 100\%}, \textit{Recall = 85.42\%}, and \textit{F1 = 92.13\%}.

\noindent \textbf{Analysis}: The 100\% precision confirms the effectiveness of new technical recognizers when properly configured. The 7 FN derived from a credential in non-conventional format and geographic locations in X.509 certificates (\texttt{DE}, \texttt{Osnabrueck}), which the NER model does not recognize in technical context. The 41 TP include IPs, \textit{hostnames}, \textit{hashes}, and complete X.509 certificates, indicating good overall performance; remaining failures concentrate on traditional PII that require contextual improvement.


\section{Discussion}
\label{sec:discussao}

\noindent \textbf{Design decisions}. The v2.0 architecture was designed to enable secure use of complex data in automated analyses and LLM training. The use of HMAC-SHA256 with \texttt{SECRET\_KEY} enables federated correlation between trusted CSIRTs without PII exposure, eliminating vulnerabilities associated with public hashes. The \texttt{--slug-length} parameter (default 64) reduces collision probability to negligible values, ensuring cryptographic integrity even in massive \textit{datasets}, while the complete hash stored in the database ensures precise re-identification. The observed variation in metrics (F1 = 76.5\% in OCR scenario and F1 = 92.13\% in OpenVAS technical scenario) reflects differences inherent to \textit{dataset} characteristics, expected behavior in NER systems.

\noindent \textbf{Limitations and future work}. Current main limitations include: (1) support for a single language per document, harming multilingual cases; and (2) dependence on manual SQLite database inspection for fine-tuning configuration, unsuitable for high-scale environments.

\textit{Future work}: (1) enhanced federated security via PKI-based key distribution protocol; (2) language detection at block/sentence level for dynamic NER model selection; (3) automated assistance with local SLM (Ollama) to analyze \texttt{entities.db}, suggest items for \texttt{--allow-list}, and generate audit reports; and (4) domain specialization through \textit{fine-tuning} of cybersecurity-specific NER models, aiming for better contextual recognition (non-standard credentials, locations in certificates, and \textit{hostnames} in terminal prompts).


\section{Final Remarks}
\label{sec:conclusao}

This paper presented AnonLFI 2.0, a complete reengineering that converts the previous version into a modular and extensible \textit{framework}, incorporating HMAC-SHA256 for cryptographic security and federated correlation, configurable slugs via \texttt{--slug-length} to eliminate collisions, specialized recognizers for technical entities, native processors that preserve hierarchical structures (JSON/XML/XLSX), OCR \textit{pipeline} for PII extraction in images, controlled re-identification CLI with audit, and multilingual support for 24 languages. Case studies with a PDF containing images (Precision = 100\%, F1 = 76.5\%) and an OpenVAS vulnerability XML (Precision = 100\%, F1 = 92.13\%) demonstrated the effectiveness of new capabilities and expansion of the original recognition engine (Microsoft Presidio, spaCy, \texttt{xlm-roberta-base-ner-hrl}), including unprecedented functionalities such as OCR and technical entity recognition. Thus, AnonLFI 2.0 enables CSIRTs to prepare complex datasets of incidents and vulnerabilities for advanced analyses and secure sharing, preserving structural integrity, cryptographic security, and regulatory compliance.

\bibliographystyle{IEEEtran}
\bibliography{sbc-template}

\end{document}